\documentclass[12pt]{article}

\usepackage{epsfig}

\usepackage{bbm}
\usepackage{amssymb,amsmath}

\newcommand{\be}{\begin{equation}}
\newcommand{\ee}{\end{equation}}
\newcommand{\ba}{\begin{eqnarray}}
\newcommand{\ea}{\end{eqnarray}}
\newcommand{\no}{\nonumber\\}

\newcommand{\mnu}{\mathcal{M}_\nu}

\newcommand{\dsol}{\Delta m^2_\odot}
\newcommand{\datm}{\Delta m^2_\mathrm{atm}}
\newcommand{\zz}{\mathbbm{Z}_2}

\newcommand{\one}{\mathbbm{1}}
\renewcommand{\S}{s_{13}^2}
\newcommand{\cs}{\mathcal{S}}
\newcommand{\xux}{\left| U_{e3} \right|}

\begin{document}

\title{\normalsize \hfill UWThPh-2008-16 \\[8mm]
\LARGE A three-parameter neutrino mass matrix \\
with maximal $CP$ violation}

\author{
W.~Grimus$^{(1)}$\thanks{E-mail: walter.grimus@univie.ac.at}
\ and
L.~Lavoura$^{(2)}$\thanks{E-mail: balio@cftp.ist.utl.pt}
\\*[3mm]
$^{(1)}$ \small
University of Vienna, Faculty of Physics \\
\small
Boltzmanngasse 5, A--1090 Vienna, Austria
\\*[2mm]
$^{(2)}$ \small
Technical University of Lisbon,
Centre for Theoretical Particle Physics \\
\small Instituto Superior T\'ecnico, 1049-001 Lisbon, Portugal
}

\date{24 October 2008}

\maketitle

\begin{abstract}
Using the seesaw mechanism,
we construct a model for the light-neutrino Majorana mass matrix
which yields trimaximal lepton mixing
together with maximal $CP$ violation
and maximal atmospheric-neutrino mixing.
We demonstrate that,
in our model,
the light-neutrino mass matrix retains its form
under the one-loop renormalization-group evolution.
With our neutrino mass matrix,
the absolute neutrino mass scale
is a function of $\xux$
and of the atmospheric mass-squared difference.
We study the effective mass in neutrinoless $\beta\beta$ decay
as a function of $\xux$,
showing that it contains a fourfold ambiguity.
\end{abstract}

\newpage

\section{Introduction}

The specific features of lepton mixing~\cite{reviews,fogli-schwetz} 
inspire the search for lepton mass matrices
which might reflect some of those features in a natural way.
In our previous paper~\cite{trimax}
we have constructed a model predicting trimaximal lepton mixing.
This is defined by $\left| U_{e2} \right|^2
= \left| U_{\mu 2} \right|^2 = \left| U_{\tau 2} \right|^2 = 1/3$,
where $U$ is the lepton mixing matrix.
Trimaximal mixing is a less stringent requirement
than tri-bimaximal mixing~\cite{HPS},
which furthermore imposes
$\left| U_{\mu 3} \right| = \left| U_{\tau 3} \right|$
and $U_{e3} = 0$.
In this paper we shall further constrain our model of~\cite{trimax}
by adding to it the non-standard $CP$ symmetry
first introduced in~\cite{nsCP}.\footnote{Earlier,
that symmetry had already been proposed in~\cite{HS-CP},
but only for neutrinos and not within a full model.}

Let us list the steps in the construction of the model and,
simultaneously,
illustrate how we proceed to reduce the number of parameters
in the lepton sector:
\begin{itemize}
\item Neutrinos are assumed to be of the Majorana type.
The smallness of the neutrino masses
is explained through the (type I) seesaw mechanism~\cite{seesaw}. 
\item Family symmetries---the discrete group $\Delta(27)$
together with three $\zz$ groups---are imposed,
and afterwards broken softly and spontaneously in a sophisticated way.
These symmetries justify specific forms
for the charged-lepton mass matrix $M_\ell$,
which is diagonal,
and for the neutrino Dirac mass matrix $M_D$,
which is proportional to the unit matrix.
The family symmetries also reduce
the number of Yukawa couplings to a minimum
and strongly constrain the Majorana mass matrix
of the right-handed neutrinos $M_R$.
Since $M_\ell$ and $M_D$ are both diagonal,
lepton mixing originates solely in $M_R$.
\item We get rid of two phases
by assuming invariance of the Lagrangian
under the non-standard $CP$ transformation of~\cite{nsCP},
which is eventually broken spontaneously at the electroweak scale.
\end{itemize}
The specific construction of the model
will be explained in the next section;
we anticipate that it produces the three-parameter
light-neutrino mass matrix~\cite{trimax}
\be
\label{mnu}
\mnu = \left( \begin{array}{ccc}
x + y & z + \omega^2 y &  z + \omega y \\
z + \omega^2 y & x + \omega y & z + y \\
z + \omega y & z + y & x + \omega^2 y
\end{array} \right),
\quad \mbox{with} \
\omega = e^{2\pi i/3}
\ \mbox{and} \
x, y, z \in \mathbbm{R},
\ee
in the basis where $M_\ell$ is diagonal.
Our model thus predicts the neutrino masses and lepton mixing---a total
of nine observables---in terms of just
three parameters---the real numbers $x$,
$y$,
and $z$.
The contribution of the non-standard $CP$ transformation
to the form of $\mnu$ is to constrain these numbers to be real;
in our original model~\cite{trimax} they could be complex and,
therefore,
the number of physical parameters in $\mnu$ was five.
 
With the $\mnu$ of equation~(\ref{mnu}),
trimaximal mixing is realized since
\be
\label{trim}
\mnu \left( \begin{array}{c} 1 \\ 1 \\ 1 \end{array} \right) =
\lambda \left( \begin{array}{c} 1 \\ 1 \\ 1 \end{array} \right),
\ee
where $\lambda = x + 2 z$.
Furthermore,
as a consequence of the non-standard $CP$ transformation,
$\mnu$ fulfills~\cite{nsCP}
\be
\label{CPrelation}
\cs \mnu \, \cs = \mnu^\ast,
\quad \mbox{where} \ \cs = \left( \begin{array}{ccc}
1 & 0 & 0 \\ 0 & 0 & 1 \\ 0 & 1 & 0
\end{array} \right).
\ee

Let us discuss the predictions of equation~(\ref{mnu})
for the neutrino masses and for lepton mixing.
We follow the convention in the Review of Particle Physics~\cite{RPP}
for the parameterization of the lepton mixing matrix.
As shown in~\cite{nsCP},
it follows from equation~(\ref{CPrelation}) that
\ba
s_{23} &=& \frac{1}{\sqrt{2}},
\\
e^{i\delta} &=& \pm i,
\ea
\textit{i.e.}\ maximal atmospheric-neutrino mixing
and maximal $CP$ violation.
Moreover,
there are no $CP$-violating Majorana phases,
\textit{i.e.}\ the values of those phases are trivial
($0$ or $\pi$). 
On the other hand,
it follows~\cite{trimax} from equation~(\ref{trim}) that
\be
\label{solmix}
s_{12}^2 = \frac{1}{3 \left( 1 - \S \right)}, 
\ee
which relates $s_{12}$ with $s_{13}$
and provides the lower limit 
\be
\label{lowerlimit}
s_{12}^2 \geq \frac{1}{3}
\ee
on $s_{12}$.
The mass matrix~(\ref{mnu}) does not determine
the type of neutrino mass spectrum;
it could be normal--- wherein the smallest neutrino mass
$m_s = m_1$---or inverted,
\textit{i.e.}\ $m_s = m_3$.
Still,
equation~(\ref{mnu}) does fix the absolute neutrino mass scale
as a function of both $\datm = \left| m_3^2 - m_1^2 \right|$ and $\S$;
indeed,
\be
\label{ms}
m_s + \sqrt{m_s^2 + \datm} =
\left[ \frac{\left( \datm \right)^2}
{3 \S \left( 2 - 3 \S \right)} \right]^{1/4}.
\ee
(This is a result from~\cite{trimax} specialized to $s_{23}^2 = 1/2$.)
In summary,
due to $\mnu$ containing only three parameters,
there are in our model six predictions
for the nine physical observables following from $\mnu$:
\begin{itemize}
\item the two Majorana phases,
the Dirac phase $\delta$,
and the atmospheric mixing angle are all fixed;
\item the solar-neutrino mixing angle is a function of $\S$
through equation~(\ref{solmix});
\item the smallest mass $m_s$ is a function of both $\S$ and $\datm$
through equation~(\ref{ms}).
\end{itemize}
In that sense, 
the parameters $x$,
$y$,
and $z$ in $\mnu$ can be traded for $\S$,
$\datm$,
and $\dsol \equiv m_2^2 - m_1^2$.

We want to stress that the $\mnu$ of equation~(\ref{mnu})
is different from the mass matrices
based on $\mu$--$\tau$ interchange symmetry.
With that symmetry one obtains $s_{13} = 0$,
therefore $CP$ violation in neutrino oscillations
is absent.\footnote{For a model with a three-parameter neutrino mass matrix 
based on $\mu$--$\tau$ interchange symmetry,
see~\cite{3-para}.}
In the model of this paper,
on the other hand,
$s_{13}$ does not need to vanish
and $CP$ violation in neutrino oscillations is maximal.
Recent fits to the oscillation data~\cite{fogli-schwetz}
indicate the possibility that $\S > 0$ at the low scale;
this would,
if confirmed,
disfavour a $\mu$--$\tau$-symmetric $\mnu$.

We also want to stress that the most general $\mnu$
satisfying equation~(\ref{trim}) is
\be
\label{trim2}
\mnu = \left( \begin{array}{ccc}
r + s & u & t \\
u & r + t & s \\
t & s & r + u
\end{array} \right),
\quad \mbox{with} \ r, s, t, u \in \mathbbm{C}
\ee
and $\lambda = r + s + t + u$.
The $\mnu$ of equation~(\ref{trim2}) contains seven parameters:
the moduli of $r$,
$s$,
$t$,
$u$ and their relative phases.
Our light-neutrino mass matrix~(\ref{mnu})
is clearly a much restricted version of~(\ref{trim2}).
Notice that,
with the $\mnu$ of equation~(\ref{trim2}),
tri-bimaximal lepton mixing is attained when $t = u$.

This paper is organized as follows.
In section~\ref{model} we discuss the model of~\cite{trimax}
under the constraint of the non-standard $CP$ symmetry of~\cite{nsCP},
demonstrating that the neutrino mass matrix~(\ref{mnu}) is obtained.
In section~\ref{renormalization-group evolution}
we study the effect of the one-loop
renormalization-group evolution on $\mnu$ and show that,
in our model,
the form~(\ref{mnu}) of $\mnu$
is not changed by that evolution,
so that the ensuing predictions hold irrespective of the energy scale.
The computation of the effective mass for neutrinoless $\beta\beta$ decay
is the subject matter of section~\ref{effective mass}.
We present our conclusions in section~\ref{concl}.

\section{The model}
\label{model}

The model that we have put forward in~\cite{trimax} 
is an extension of the Standard Model,
with gauge group $SU(2) \times U(1)$.
The lepton multiplets are the standard left-handed $SU(2)$ doublets
$D_{\alpha L} = \left( \nu_{\alpha L}, \, \alpha_L \right)^T$,
the right-handed charged-lepton $SU(2)$ singlets $\alpha_R$,
and four right-handed-neutrino $SU(2)$ singlets $\nu_{\alpha R}$,
$\nu_{0R}$
($\alpha = e, \mu, \tau$).
The scalar sector of the Standard Model is also extended
to four $SU(2)$ doublets $\phi_\alpha$,
$\phi_0$,
together with a complex gauge singlet $S$.

Next we discuss the symmetries of the model.
The two $3 \times 3$ matrices 
\be\label{FT}
F = \left( \begin{array}{ccc}
0 & 0 & 1 \\ 1 & 0 & 0 \\ 0 & 1 & 0
\end{array} \right),
\quad
T = \left( \begin{array}{ccc}
1 & 0 & 0 \\ 0 & \omega & 0 \\ 0 & 0 & \omega^2
\end{array} \right)
\ee
satisfy $F^3 = T^3 = \one$ and do not commute.
Together they generate one of the two
three-dimensional faithful irreducible representations
of the discrete group $\Delta(27)$~\cite{luhn};
the other one is generated by $F$ and $T^\ast$.
We display in table~\ref{transformationCT}
the way in which the multiplets with index $\alpha$
transform under this horizontal symmetry $\Delta(27)$.
\begin{table}
\begin{center}
\begin{tabular}{c|cccc}
& $D_{\alpha L}$ & $\alpha_R$ & $\nu_{\alpha R}$ & $\phi_\alpha$ \\ \hline
$F$ & $F$ & $F$ & $F$ & $F$ \\
$T$ & $T$ & $T^\ast$ & $T$ & $T^\ast$ 
\end{tabular}
\end{center}
\caption{Transformation properties of the $\Delta(27)$ triplets
under $F$ and $T$.
\label{transformationCT}}
\end{table}
The multiplets without index $\alpha$ are $T$-invariant
and transform as
\be
\nu_{0R} \to \omega \nu_{0R},
\quad
\phi_0 \to \phi_0,
\quad
S \to \omega S
\quad \mbox{under} \ F.
\ee
Next we introduce three $\zz$ symmetries 
$\mathbf{z}_{e, \mu, \tau}$ as~\cite{GL06} 
\be
\label{z}
\mathbf{z}_\alpha: \quad 
\alpha_R \to - \alpha_R, \
\phi_\alpha \to - \phi_\alpha,
\ee
and all other multiplets remain unchanged.
Finally,
we come to the non-standard $CP$ transformation~\cite{nsCP},
defined by
\be
\label{CPtrafof}
D_{\alpha L}   \to i \mathcal{S}_{\alpha\beta}
\gamma_0 C \bar D_{\beta L}^T,
\quad
\alpha_R \to i \mathcal{S}_{\alpha\beta}
\gamma_0 C \bar\beta_R^T,
\quad
\nu_{\alpha R} \to i \mathcal{S}_{\alpha\beta}
\gamma_0 C \bar \nu_{\beta R}^T,
\quad
\nu_{0 R} \to i \gamma_0 C \bar \nu_{0 R}^T,
\ee
for the fermions ($C$ is the charge-conjugation matrix) and
\be
\label{CPtrafos}
\phi_\alpha \to \mathcal{S}_{\alpha\beta} \phi_\beta^\ast,
\quad
\phi_0 \to \phi_0^\ast,
\quad
S \to S^\ast
\ee
for the scalars.
The matrix $\cs$ is given in equation~(\ref{CPrelation}).

The multiplets and symmetries lead to the Yukawa couplings~\cite{trimax}
\be
\mathcal{L}_Y =
- y_1 \sum_{\alpha = e, \mu, \tau}
\bar D_{\alpha L} \alpha_R \phi_\alpha
\,-\, y_4 \sum_{\alpha = e, \mu, \tau}
\bar D_{\alpha L} \nu_{ \alpha R}
\left( i \tau_2 \phi_0^\ast \right)
\,+\, \frac{y_5}{2}\, \nu_{0R}^T C^{-1} \nu_{0R} \, S
+ \mathrm{H.c.}
\label{LY}
\ee
In the Majorana mass terms we allow
soft breaking of the symmetry $T$,
but not of $F$.
Therefore,~\cite{trimax}
\ba
\mathcal{L}_\mathrm{Maj} &=&
\frac{M_0}{2} \sum_{\alpha = e, \mu, \tau}
\nu_{\alpha R}^T C^{-1} \nu_{\alpha R}
+ M_1 \left(
\nu_{eR}^T C^{-1} \nu_{\mu R}
+ \nu_{\mu R}^T C^{-1} \nu_{\tau R}
+ \nu_{\tau R}^T C^{-1} \nu_{eR}
\right)
\no & &
+ \frac{M_2}{2} \left(
\nu_{eR}^T + \omega \nu_{\mu R}^T + \omega^2 \nu_{\tau R}^T
\right) C^{-1} \nu_{0R} + \mathrm{H.c.}
\label{LM}
\ea
Note the consequences of the non-standard $CP$ symmetry:
\be
y_1, y_4, y_5, M_0, M_1, M_2 \in \mathbbm{R}.
\ee

Since the first term in $\mathcal{L}_Y$
has one common coupling constant $y_1$,
upon spontaneous symmetry breaking
one needs three different vacuum expectation values (VEVs)
$v_\alpha = \left\langle \phi_\alpha^0 \right\rangle_0$ 
to be able to account for the different charged-lepton masses:
\be
\label{ccm}
m_e : m_\mu : m_\tau
= \left| v_e \right| : \left| v_\mu \right| : \left| v_\tau \right|.
\ee

From equations~(\ref{LY}) and~(\ref{LM}),
we find the Majorana and Dirac neutrino mass matrices
\be
\label{MRD}
M_R = \left( \begin{array}{cccc}
M_0 & M_1 & M_1 & M_2 \\
M_1 & M_0 & M_1 & \omega^2 M_2 \\
M_1 & M_1 & M_0 & \omega M_2 \\
M_2 & \omega^2 M_2 & \omega M_2 & M_N
\end{array} \right),
\quad
M_D = \left( \begin{array}{ccc}
a & 0 & 0 \\ 0 & a & 0 \\ 0 & 0 & a \\ 0 & 0 & 0 
\end{array} \right),
\ee
respectively.
We have defined $M_N = y_5 v_S^\ast$ and $a = y_4 v_0$,
where $v_S$ is the VEV of $S$
and $v_0$ is the VEV of the neutral component of $\phi_0$.

With $\mnu = - M_D^T M_R^{-1} M_D$,
the mass matrix~(\ref{mnu}) is obtained.
Since $M_{0,1,2} \in \mathbbm{R}$,
it follows that the parameters $x$ and $z$ of $\mnu$
are real as well---see~\cite{trimax}.
On the other hand,
$y \propto 1/M_N$~\cite{trimax};
therefore,
the question whether $y$ is real
boils down to the question
whether the minimum of the scalar potential $V$ occurs for a real $v_S$.
For this reason,
we next investigate $V$.

The terms of dimension four in $V$
must be invariant under $F$,
$T$,
and $\mathbf{z}_{e, \mu, \tau}$.
Terms of dimension three are invariant
under $F$ and  $\mathbf{z}_{e, \mu, \tau}$,
but they may violate $T$,
\textit{cf.}~equation~(\ref{LM}).
If one requires terms of dimension two
to be invariant under $\mathbf{z}_{e, \mu, \tau}$ as well,
then there are four $U(1)$ symmetries in the potential,
one for each Higgs doublet;
these symmetries are all spontaneously broken,
resulting in three physical Goldstone bosons.
To avoid this problem we break
the $\mathbf{z}_\alpha$ softly by terms of dimension two.
As for the symmetry $F$,
we want it to be spontaneously broken
through different VEVs $v_{e, \mu, \tau}$;
in order to guarantee that this happens,
we must admit $F$ to be softly broken by terms of dimension two too.
The only symmetry respected by the full scalar potential
is the $CP$ symmetry.
Thus,
\ba
V &=& 
\lambda_1 \left[
\left( \phi_e^\dagger \phi_e \right)^2
+ \left( \phi_\mu^\dagger \phi_\mu \right)^2
+ \left( \phi_\tau^\dagger \phi_\tau \right)^2 \right]
+ \lambda_2 \left( \phi_0^\dagger \phi_0 \right)^2
\no & &
+ \lambda_3 \left(
\phi_e^\dagger \phi_e \, \phi_\mu^\dagger \phi_\mu
+ \phi_\mu^\dagger \phi_\mu \, \phi_\tau^\dagger \phi_\tau
+  \phi_\tau^\dagger \phi_\tau \, \phi_e^\dagger \phi_e \right)
+ \lambda_4 \, \phi_0^\dagger \phi_0
\left( \phi_e^\dagger \phi_e + \phi_\mu^\dagger \phi_\mu
+ \phi_\tau^\dagger \phi_\tau \right)
\no & &
+ \lambda_5 \left(
\phi_e^\dagger \phi_\mu \, \phi_\mu^\dagger \phi_e
+ \phi_\mu^\dagger \phi_\tau \, \phi_\tau^\dagger \phi_\mu
+ \phi_\tau^\dagger \phi_e \, \phi_e^\dagger \phi_\tau \right)
+ \lambda_6 \, \phi_0^\dagger \left( 
\phi_e \phi_e^\dagger + \phi_\mu \phi_\mu^\dagger
+ \phi_\tau \phi_\tau^\dagger \right) \phi_0
\no & &
+ \left| S \right|^2 
\left[ \lambda_7 \left( \phi_e^\dagger \phi_e
+ \phi_\mu^\dagger \phi_\mu + \phi_\tau^\dagger \phi_\tau \right)
+ \lambda_8 \, \phi_0^\dagger \phi_0 \right]
\no & &
+ \lambda_9 \left[
S^2 \left( \phi_e^\dagger \phi_e
+ \omega^2 \phi_\mu^\dagger \phi_\mu
+ \omega \phi_\tau^\dagger \phi_\tau \right)
+ {S^2}^\ast \left( \phi_e^\dagger \phi_e
+ \omega \phi_\mu^\dagger \phi_\mu
+ \omega^2 \phi_\tau^\dagger \phi_\tau \right) \right]
\no & &
+ \bar\mu \left[
S \left( \phi_e^\dagger \phi_e
+ \omega \phi_\mu^\dagger \phi_\mu
+ \omega^2 \phi_\tau^\dagger \phi_\tau \right)
+ S^\ast \left( \phi_e^\dagger \phi_e
+ \omega^2 \phi_\mu^\dagger \phi_\mu
+ \omega \phi_\tau^\dagger \phi_\tau \right) \right]
\no & &
+ \mu \left| S \right|^2
+ \lambda \left| S \right|^4
+ \mu_1 \left( S^3 + {S^\ast}^3 \right)
+ \mu_2 \left( S^2 + {S^\ast}^2 \right)
+ \mu_3 \left( S + S^\ast \right)
\no & &
+ \sum_{i,j=0,e,\mu,\tau} 
\left( \mathcal{M}_\phi \right)_{ij} \phi_i^\dagger \phi_j.
\label{V}
\ea
All the parameters in $V$ are real;
the only exception
are the parameters of the $4 \times 4$ matrix $\mathcal{M}_\phi$,
which has the structure
\be
\mathcal{M}_\phi = \left( \begin{array}{cccc}
a & b & c & c^\ast \\
b & d & e & e^\ast \\
c^\ast & e^\ast & f & g \\
c & e & g^\ast & f
\end{array} \right),
\quad \mbox{with} \ a, b, d, f \in \mathbbm{R},
\ee
in the basis $\left( \phi_0, \phi_e, \phi_\mu, \phi_\tau \right)$.
Note that $c$,
$e$,
and $g$ are in general complex.
These conditions derive from hermiticity together with $CP$ invariance.
Note that $\mathcal{M}_\phi$ is still general enough
to allow for different VEVs $v_i$ ($i = 0, e, \mu, \tau$).

Since the VEVs $v_i$ are of the electroweak scale,
whereas $v_S$ is of the seesaw scale,
a fine-tuning is necessary in $V$,
with extremely small coupling constants $\lambda_7$,
$\lambda_8$,
$\lambda_9$,
and $\bar\mu$.
This is an unpleasant feature of the model,
which is shared,
though,
by all other models with two widely different mass scales.

Now we address the conditions for obtaining a real VEV of $S$.
For this purpose we only have to consider
the penultimate line of equation~(\ref{V}).
Obviously,
if
\be
\mbox{sign}\, \mu_1
= \mbox{sign}\, \mu_3
\quad \mbox{and} \ \mu_2 < 0,
\ee
then the absolute minimum of $V$ is attained for real  $v_S$ 
($v_S > 0$ for $\mu_1 < 0$ and $v_S < 0$ for $\mu_1 > 0$).

We find it useful to summarize the symmetry breaking in our model:
\begin{itemize}
\item At the seesaw scale,
$T$ is softly broken
by terms of dimension three in $\mathcal{L}_\mathrm{Maj}$,
while $F$ is spontaneously broken
by the VEV of $S$.\footnote{Notice that the VEV of $S$
is crucial in our model in order to obtain $M_N \neq 0$.
If it were not for $M_N \neq 0$,
the seesaw mechanism would be unable to suppress the masses
of all three light neutrinos,
because $\det{M_R} \propto M_N$.}
\item At the electroweak scale,
$T$,
$F$,
and the $\mathbf{z}_\alpha$ are all softly broken
in the matrix $\mathcal{M}_\phi$.
All symmetries,
including the non-standard $CP$ symmetry,
are broken spontaneously at this scale.
\end{itemize}
The $CP$ symmetry is broken by 
$\left| v_\mu \right| \neq \left| v_\tau \right|$;
the different VEVs are necessary in our model for having
$m_\mu \neq m_\tau$---see equation~(\ref{ccm}).
Now we want to show that indeed 
there is no violation of $CP$ 
for degenerate masses $m_\mu = m_\tau$.
In that case we may transform
\be
M_R \to K M_R K^T \quad \mbox{with} \
K = \left( \begin{array}{ccc} 1 & 0_{1 \times 2} & 0 \\
0_{2 \times 1} & K^\prime & 0_{2 \times 1} \\
0 & 0_{1 \times 2} & 1
\end{array} \right),
\ee
where $K^\prime$ is a $2 \times 2$ unitary matrix.
Choosing
\be
K^\prime = \frac{1}{\sqrt{2}}
\left( \begin{array}{cc} i & -i \\ 1 & 1 \end{array} \right),
\ee
one sees that
\be
M_R \to \left( \begin{array}{cccc}
M_0 & 0 & \sqrt{2} M_1 & M_2 \\
0 & M_1 - M_0 & 0 & \sqrt{3} M_2 \left/ \sqrt{2} \right. \\
\sqrt{2} M_1 & 0 & M_1 + M_0 & - M_2 \left/ \sqrt{2} \right. \\
M_2 & \sqrt{3} M_2 \left/ \sqrt{2} \right. & - M_2 \left/ \sqrt{2} \right.
& M_N
\end{array} \right)
\ee
becomes real.
Thus,
it is indeed $\left| v_ \mu \right| \neq \left| v_\tau \right|$
which breaks the $CP$ symmetry,
at the electroweak scale.
Notice that the phases of the VEVs are irrelevant
in the breaking of $CP$.

\section{The renormalization-group evolution}
\label{renormalization-group evolution}

Throughout this section,
we use indices $i, j, \ldots = 0, e, \mu, \tau$
to refer to the four Higgs doublets of our model.

We use the formalism of our paper~\cite{renorm}.
In that paper we have considered
a multi-Higgs-doublet extension of the Standard Model
supplemented by effective dimension-five operators
\be
\label{operators}
\mathcal{O}_{ij} = \sum_{\alpha, \beta = e, \mu, \tau}
\kappa^{(ij)}_{\alpha \beta}
\left( \nu_{\alpha L}^T \phi_i^0 - \alpha_L^T \phi_i^+ \right)
C^{-1}
\left( \nu_{\beta L} \phi_j^0 - \beta_L \phi_j^+ \right),
\ee
where the $\kappa^{(ij)}_{\alpha \beta}$
are coefficients with dimension $-1$.
This is sufficient for our purposes despite the occurrence,
in our model,
of a scalar singlet $S$.
The reason is that $S$ is integrated out at the seesaw scale $m_R$,
because $S$ has VEV and mass of order $m_R$.

In our model there are symmetries $F$,
$T$,
and $\mathbf{z}_\alpha$.\footnote{There is also the $CP$ symmetry.
The reasoning and the conclusions in this section are,
however,
independent of the presence or absence of that $CP$ symmetry.
Thus,
the present section and its conclusions apply as well
to the model of~\cite{trimax}.}
As emphasized at the end of the last section,
both $F$ and $T$ are broken at the seesaw scale $m_R$,
while the three symmetries $\mathbf{z}_\alpha$
stay valid below $m_R$ and are broken,
both softly and spontaneously,
only at the electroweak scale $m_F$.
Thus,
at the relevant energy scales,
\textit{i.e.}~in between $m_R$ and $m_F$,
the symmetries $\mathbf{z}_\alpha$ hold.
We therefrom conclude that in our model
the only non-zero matrices $\kappa^{(ij)}$ are those with $i=j$.

Upon spontaneous symmetry breaking
the light-neutrino mass terms are
\be
\frac{1}{2} \sum_ {\alpha, \beta} \left( \mnu \right)_{\alpha \beta}
\nu_{L \alpha}^T C^{-1} \nu_{L \beta}
\quad \mbox{with} \
\frac{\mnu}{2} = \sum_{i, j} v_i v_j \kappa^{(ij)},
\ee
where $v_i$ is the VEV of $\phi_i^0$.
Since only the $\kappa^{(ii)}$ are non-zero,
we have,
at any energy scale,
\be
\label{mnusum}
\mnu = 2 \sum_{i} v_i^2 \kappa^{(ii)}.
\ee
The matrices $\kappa^{(ii)}$ are symmetric,
as is obvious from equation~(\ref{operators}).

At the high scale $m_R$,
only the Higgs doublet $\phi_0$
(and also the scalar singlet $S$)
has Yukawa couplings to the right-handed neutrinos,
while the other three Higgs doublets $\phi_\alpha$ have no such couplings.
We therefrom conclude that,
at the scale $m_R$,
\be
\label{initialk}
\kappa^{(00)} = \frac{\mnu}{2 v_0^2},
\quad
\kappa^{(ii)} = 0_{3 \times 3}
\ \mbox{for} \ i = e, \mu, \tau.
\ee
This is the initial condition for the renormalization-group (RG) running,
with $\mnu$ given by equation~(\ref{mnu}).

Let us write the scalar potential of the Higgs doublets in the form
\be
V = \mbox{quadratic \ terms} + \sum_{i,j,k,l} \lambda_{ijkl}\,
\phi_i^\dagger \phi_j \, \phi_k^\dagger \phi_l.
\ee
Let us furthermore write the Yukawa couplings of the Higgs doublets
to the right-handed charged leptons in the form
\be
\mathcal{L}_{Y \ell} = - \sum_i \sum_{\alpha, \beta}
\left( Y_i \right)_{\alpha \beta}
\bar D_{\alpha L} \beta_R \phi_i
+ \mbox{H.c.}
\ee
Then,
the RG equations for the evolution of the matrices $\kappa^{(ii)}$
are~\cite{renorm,r1}
\ba
16 \pi^2\, \frac{\mathrm{d} \kappa^{(ii)}}{\mathrm{d} t} &=&
\left( - 3 g^2 + 2 T_{ii} \right) \kappa^{(ii)}
+ \kappa^{(ii)} P + P^T \kappa^{(ii)}
- 2 \left[ \kappa^{(ii)} Y_i Y_i^\dagger
+ Y_i^\ast Y_i^T \kappa^{(ii)} \right]
\no & &
+ 4 \sum_j \lambda_{jiji} \kappa^{(jj)}.
\ea
Here,
$t = \ln{\mu}$ is the logarithm of the mass scale,
$g$ is the $SU(2)$ gauge coupling constant,
and
\ba
T_{jk} &=& \mathrm{tr} \left( Y_j^\dagger Y_k \right),
\\
P &=& \frac{1}{2} \sum_j Y_j Y_j^\dagger.
\ea
If we define the $3 \times 3$ matrices
\be
P_0 = 0_{3 \times 3}, \quad 
P_e = \mbox{diag} \left( 1, 0, 0 \right), \quad
P_\mu = \mbox{diag} \left( 0, 1, 0 \right), \quad
P_\tau = \mbox{diag} \left( 0, 0, 1 \right),
\ee
then we see in equation~(\ref{LY}) that,
in our model,
at the scale $m_R$, 
\be
Y_i = y_1 P_i \quad \mbox{for} \ i = 0, e, \mu, \tau,
\label{yi}
\ee
and,
therefore,
\be
T_{ee} = T_{\mu \mu} = T_{\tau \tau} = \left| y_1 \right|^2
\
\mathrm{and \ all \ other} \ T_{ij} = 0,
\quad
P = \frac{\left| y_1 \right|^2}{2}\, \mathbbm{1}_ {3 \times 3}.
\ee
Now one can easily check that the matrices $Y_i$
remain of the form~(\ref{yi}) at all energies
when they evolve with the RG equations
for the Yukawa couplings.
We therefore have
\ba
16 \pi^2\, \frac{\mathrm{d} \kappa^{(00)}}{\mathrm{d} t} &=&
\left( - 3 g^2 + \left| y_1 \right|^2 \right) \kappa^{(00)}
+ 4 \sum_j \lambda_{j0j0} \kappa^{(jj)},
\label{00}
\\
16 \pi^2\, \frac{\mathrm{d} \kappa^{(\alpha \alpha)}}{\mathrm{d} t} &=&
\left( - 3 g^2 + 3 \left| y_1 \right|^2 \right) \kappa^{(\alpha \alpha)}
- 2 \left| y_1 \right|^2
\left\{ \kappa^{(\alpha \alpha)}, P_\alpha \right\}
+ 4 \sum_j \lambda_{j \alpha j \alpha} \kappa^{(jj)},
\label{aa}
\ea
where the anti-commutator of matrices $A$ and $B$
is denoted $\left\{ A, B \right\}$.

We observe that the coefficients $\lambda_{jiji}$
are particularly important in equations~(\ref{00}) and~(\ref{aa}).
According to~\cite{renorm},
the RG equation for those coefficients is
\ba
\label{lr}
16 \pi^2\, \frac{\mathrm{d} \lambda_{ijij}}{\mathrm{d} t} &=&
4 \sum_{k,l} \left( 2 \lambda_{ijkl} \lambda_{lkij}
+ \lambda_{ijkl} \lambda_{iklj}
+ \lambda_{iklj} \lambda_{klij}
+ \lambda_{ikil} \lambda_{kjlj}
+ \lambda_{kjil} \lambda_{iklj} \right)
\no & &
- \left( 9 g^2 + 3 {g'}^2 - 2 T_{ii} - 2 T_{jj} \right) \lambda_{ijij}
\ea
whenever $i \neq j$.
In equation~(\ref{lr}),
$g^\prime$ is the $U(1)$ gauge coupling constant.
A little contemplation of equation~(\ref{lr}) 
allows one to ascertain that,
if all the $\lambda_{ijij}$ with $i \neq j$ vanish at the scale $m_R$,
then no such non-vanishing coefficient
will ever arise through the RG evolution in our model.

Now,
this is precisely the situation that occurs.
As seen in equation~(\ref{V}),
at the scale $m_R$,
\textit{i.e.}~at the initial condition for the RG evolution,
there is no term $\left( \phi_i^\dagger \phi_j \right)^2$
with $i \neq j$ in the scalar potential.
Therefore,
all coefficients $\lambda_{ijij}$ with $i \neq j$
vanish at the scale $\mu = m_R$,
and,
through equation~(\ref{lr}),
at all other scales too.

Equations~(\ref{00}) and~(\ref{aa}),
together with the initial condition~(\ref{initialk}),
then allow us to ascertain that,
in our model,
\emph{all matrices $\kappa^{(ij)}$ except $\kappa^{(00)}$
vanish at all energy scales}.
The sole non-vanishing $\kappa^{(00)}$ matrix evolves according to
\be
16 \pi^2\, \frac{\mathrm{d} \kappa^{(00)}}{\mathrm{d} t} =
\left( - 3 g^2 + \left| y_1 \right|^2 + 4 \lambda_{0000} \right)
\kappa^{(00)},
\ee
and therefore it does not change its form along the RG evolution.
This means that
\emph{all the predictions of our model are RG-invariant}.

In summary
this result comes from the fact that
the symmetries of our model suitably constrain
both the Yukawa and the quartic Higgs couplings,
and that soft and spontaneous breaking at the seesaw scale---which is
necessary in our model for obtaining
the desired $\mnu$---has no impact on their RG equations.

\section{The effective mass in neutrinoless $\beta\beta$ decay}
\label{effective mass}

In our model the effective mass relevant for neutrinoless
$\beta \beta$ decay is
\be
m_{\beta \beta} = \left| \left( \mnu \right)_{ee} \right|
= \left| x + y \right|.
\ee
In order to compute this we have to diagonalize $\mnu$.
We define~\cite{HPS}
\be
U_\mathrm{HPS} \equiv \left( \begin{array}{ccc}
2/\sqrt{6} & 1/\sqrt{3} & 0 \\ 
-1/\sqrt{6} & 1/\sqrt{3} & -1/\sqrt{2} \\ 
-1/\sqrt{6} & 1/\sqrt{3} & 1/\sqrt{2}
\end{array} \right)
\ee
and compute
\be
\label{13}
U_\mathrm{HPS}^T \mnu U_\mathrm{HPS} = \left( \begin{array}{ccc}
p+q & 0 & i q \\ 0 & x + 2z & 0 \\ i q & 0 & p-q
\end{array} \right)
\quad \mbox{with} \ p = x-z \ \mbox{and} \ q = \frac{3 y}{2}.
\ee
Therefore,
defining
\be
\label{Vi}
V = U_\mathrm{HPS}\, D_i\, K,
\quad \mbox{with} \ D_i = \mbox{diag} \left( 1, 1, i \right)
\ \mbox{and} \ K = \left( \begin{array}{ccc}
\cos{\alpha} & 0 & \sin{\alpha} \\ 0 & 1 & 0 \\
- \sin{\alpha} & 0 & \cos{\alpha} 
\end{array} \right),
\ee
it is evident that the mass matrix $\mnu$ can be diagonalized as
\be
V^T \! \mnu V = \mbox{diag} \left( \lambda_1, \lambda_2, \lambda_3 \right),
\ee
where the neutrino masses are given by 
$m_j = \left| \lambda_j \right|$ ($j = 1, 2, 3$)
and the full diagonalization matrix  is
\be
U = V \, \mbox{diag} \left( \eta_1, \eta_2, \eta_3 \right),
\ee
where $\eta_j = 1$ for $\lambda_j > 0$
and $\eta_j = i$ for $\lambda_j < 0$.

Then,
\ba
\left( \mnu \right)_{ee} &=&
\lambda_1 \left( V_{e1}^\ast \right)^2
+ \lambda_2 \left( V_{e2}^\ast \right)^2
+ \lambda_3 \left( V_{e3}^\ast \right)^2
\no &=&
\frac{2 \lambda_1 + \lambda_2}{3} +
\left( \lambda_3 - \lambda_1 \right) \frac{2 \sin^2{\alpha}}{3}
\no &=&
\frac{2 \lambda_1 + \lambda_2}{3} +
\left( \lambda_3 - \lambda_1 \right)
\frac{1 - \cos{2 \alpha}}{3}
\no &=&
\label{uytdf}
\frac{\lambda_1 + \lambda_2 + \lambda_3}{3}
+ \frac{\left( \lambda_1 - \lambda_3 \right) \cos{2 \alpha}}{3}.
\ea

Since
\be
\mbox{diag} \left( \lambda_1, \lambda_3 \right) =
\left( \begin{array}{cc}
\cos{\alpha} & - \sin{\alpha} \\ \sin{\alpha} & \cos{\alpha}
\end{array} \right)
\left( \begin{array}{cc}
p + q & - q \\ - q & - p + q
\end{array} \right)
\left( \begin{array}{cc}
\cos{\alpha} & \sin{\alpha} \\ - \sin{\alpha} & \cos{\alpha}
\end{array} \right),
\ee
one finds that
\ba
\sin{2 \alpha} &=& \frac{-q}{\epsilon \sqrt{p^2 + q^2}},
\\
\cos{2 \alpha} &=& \frac{-p}{\epsilon \sqrt{p^2 + q^2}},
\label{utynf} \\
\lambda_1 &=& q - \epsilon \sqrt{p^2 + q^2},
\\
\lambda_3 &=& q + \epsilon \sqrt{p^2 + q^2},
\label{mhjgp}
\ea
where $\epsilon = \pm 1$.
Thus,
\be
\lambda_1 \lambda_3 = - p^2 < 0,
\ee
hence
\be
p = \eta \sqrt{- \lambda_1 \lambda_3},
\label{p}
\ee
where $\eta = \pm 1$.
From equations~(\ref{utynf})--(\ref{mhjgp}),
we obtain 
\be
\cos{2 \alpha} = \frac{-2p}{\lambda_3 - \lambda_1}.
\label{cos}
\ee
Returning to equation~(\ref{uytdf}),
one obtains
\be
\left( \mnu \right)_{ee} = \frac{\lambda_1 + \lambda_2 + \lambda_3
+ 2 \eta \sqrt{- \lambda_1 \lambda_3}}{3}.
\label{fund}
\ee

Now we must examine the matter of sign ambiguities
in $\left( \mnu \right)_{ee}$.
The three $\lambda_j$ may be either positive or negative,
but they are subject to the condition $\lambda_1 \lambda_3 < 0$.
Thus,
\be
\lambda_1 = \zeta m_1, \
\lambda_2 = \varepsilon m_2, \
\lambda_3 = - \zeta m_3, \quad
\mbox{with} \ \zeta = \pm 1 \ \mbox{and} \ \varepsilon = \pm 1.
\ee
On the other hand,
\be
\left| U_{e3} \right|^2 = \frac{2 \sin^2{\alpha}}{3}
\ee
is experimentally known to be very small,
or even zero.
Hence,
\be
\cos{2 \alpha} = 1 - 3 \left| U_{e3} \right|^2 > 0.
\ee
Therefore,
from equations~(\ref{p}) and~(\ref{cos}),
\be
\frac{- 2 \eta \sqrt{- \lambda_1 \lambda_3}}{\lambda_3 - \lambda_1} > 0,
\ee
or
\be
\frac{2 \eta \sqrt{- \lambda_1 \lambda_3}}
{\zeta \left( m_1 + m_3 \right)} > 0.
\ee
Therefore,
\be
\eta \sqrt{- \lambda_1 \lambda_3} = \zeta \sqrt{m_1 m_3}.
\ee
Thus,
from equation~(\ref{fund}),
\be
3 \left( \mnu \right)_{ee} = 
\varepsilon m_2 + \zeta \left( m_1 - m_3 + 2 \sqrt{m_1 m_3} \right).
\ee
There are therefore two possibilities:
%
%
%
\be
\label{hip}
\begin{array}{rcl}
\mbox{either} \ m_{\beta \beta} &=& \displaystyle\frac{1}{3}
\left| m_2 + m_1 - m_3 + 2 \sqrt{m_1 m_3} \right|,
\\[3mm]
\mbox{or} \ m_{\beta \beta} &=& \displaystyle\frac{1}{3}
\left| m_2 - m_1 + m_3 - 2 \sqrt{m_1 m_3} \right|.
\end{array}
\ee
Both these possibilities exist
independently of whether the neutrino mass spectrum is normal or inverted.
Thus,
besides~(\ref{hip}),
\be
\begin{array}{rcl}
&\mbox{either} \ m_1 = m_s, \
m_2 = \sqrt{m_s^2 + \dsol}, \
m_3 = \sqrt{m_s^2 + \datm}, &
\\[3mm]
&\mbox{or} \ m_1 = \sqrt{m_s^2 + \datm}, \
m_2 = \sqrt{m_s^2 + \datm + \dsol}, \
m_3 = m_s. &
\end{array}
\ee
\begin{figure}[bt]
\begin{center}
\epsfig{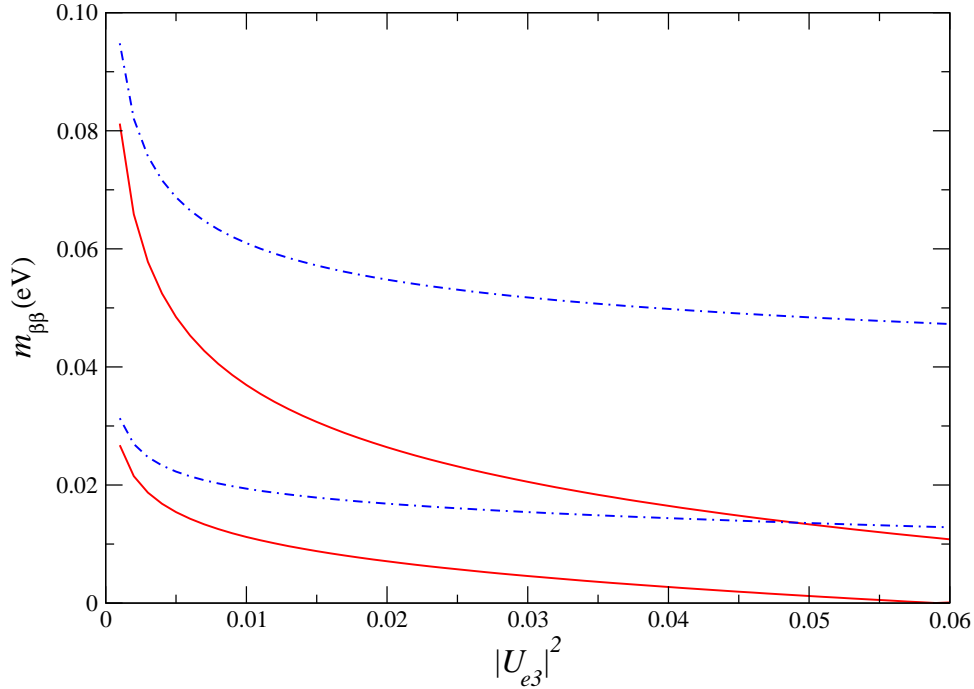}
\end{center}
\caption{The effective mass in neutrinoless $\beta\beta$ decay,
$m_{\beta\beta}$,
as a function of $\left| U_{e3} \right|^2$. 
The full lines and the dashed-dotted lines
correspond to a normal and an inverted neutrino mass spectrum,
respectively.
We have fixed the neutrino mass-squared differences
at their mean values given in the second paper of~\cite{fogli-schwetz}. 
\label{mbb-fig}}
\end{figure}
In figure~\ref{mbb-fig} we have plotted 
$m_{\beta\beta}$ as a function of $\S$, taking into account 
all four possibilities explained above. The input values for the
mass-squared differences are
$\datm = 2.40 \times 10^{-3}$ eV$^2$ and 
$\dsol = 7.65 \times 10^{-5}$ eV$^2$, 
the mean values given in the second paper of~\cite{fogli-schwetz}.

\section{Conclusions}
\label{concl}

We have presented in this paper a model whose symmetries force
the charged-lepton mass matrix to be diagonal while generating
the highly predictive three-parameter neutrino mass matrix
of~(\ref{mnu}).
It follows from this neutrino mass matrix
that the lepton mixing matrix $U$ has only one free parameter,
which may be chosen to be $s_{13}$,
while both atmospheric-neutrino mixing and $CP$ violation
are fixed and maximal.
Note that,
while $U$ displays a maximal Dirac phase,
it has vanishing Majorana phases.
Since the model leads to trimaximal mixing,
$s_{12}^2$ must be larger than $1/3$---see equations~(\ref{solmix})
and~(\ref{lowerlimit}).
This is slightly disfavoured by the present data,
but the value $1/3$ is still within the $2 \sigma$ range for $s_{12}^2$.
In any case,
the correlation~(\ref{solmix}) between $s_{12}^2$ and $s_{13}^2$
is a crucial test of trimaximal mixing.

In our model,
the symmetry which leads to maximal atmospheric-neutrino mixing
and maximal $CP$ violation
is the non-standard $CP$ transformation
given by equations~(\ref{CPtrafof}) and~(\ref{CPtrafos}).
A $CP$ symmetry of this type has the curious property that
$m_\mu \neq m_\tau$ is an effect of its spontaneous breaking,
as was noticed earlier in~\cite{nsCP,neufeld}. 

Given $\S$ and $\datm$,
the mass matrix~(\ref{mnu}) fixes the absolute neutrino mass
scale---see equation~(\ref{ms})---but
does not determine the type of neutrino mass spectrum.
Because of a sign ambiguity there are,
for each type of spectrum,
two possibilities for the effective mass $m_{\beta\beta}$
of neutrinoless $\beta\beta$ decay as a function of $\S$.
If the neutrino mass spectrum is of the inverted type,
one of the possibilities for $m_{\beta\beta}$
is within the projected range of future experiments.

We have also presented in this paper
a mechanism for the one-loop renormalization-group stability
of the neutrino mass matrix.
Indeed,
although some soft and spontaneous breaking
of symmetries
occurs already at the seesaw scale,
the symmetries of our model are such that,
in between the seesaw and the electroweak energy scales,
only the effective dimension-5 neutrino-mass operator
associated with the Higgs doublet $\phi_0$ is non-zero,
but that Higgs doublet is different
from the Higgs doublets $\phi_{e, \mu, \tau}$
which give mass to the charged leptons.
Therefore,
the form of the neutrino mass matrix is not RG-distorted
by the fact that all three charged leptons have different masses.

\paragraph{Acknowledgements:}
We acknowledge support from the European Union
through the network programme MRTN-CT-2006-035505.
The work of L.L.~was supported by the Portuguese
\textit{Funda\c c\~ao para a Ci\^encia e a Tecnologia}
through the project U777--Plurianual.

\end{document}